\documentclass[usenatbib,usegraphicx,usepdflatex]{mn2e}
\usepackage{rotating}

\newcommand{\HI}{\mbox{\sc H i}}
\newcommand{\HII}{\mbox{\sc H ii}}
\newcommand{\MHI}{\mbox{${\cal M}_{\rm HI}$}}
\newcommand{\Msun}{\mbox{${\cal M}_\odot$}}
\newcommand{\Qsg}{\mbox{$Q_{\rm 2f}$}}
\newcommand{\Zeta}{\mbox{${\cal Z}$}}
\def\lapeq{\mathrel{\hbox{\rlap{\hbox{\lower4pt\hbox{$\sim$}}}\hbox{$<$}}}}
\def\gapeq{\mathrel{\hbox{\rlap{\hbox{\lower4pt\hbox{$\sim$}}}\hbox{$>$}}}}

\title[\HI\ as a Dark Matter Tracer]{Disk stability and neutral hydrogen
  as a tracer of dark matter}
\author[G.R.\ Meurer et al.]{Gerhardt R.\ Meurer$^1$, Zheng
  Zheng$^2$, and W.J.G. de Blok$^{3,4}$\\
$^1$International Centre for Radio Astronomy Research, The University of
Western Australia, 35 Stirling Highway,\\ Crawley, WA 6009, Australia\\
$^2$The Johns Hopkins University, Department of Physics and Astronomy,
Baltimore, MD 21218, U.S.A.\\
$^3$Netherlands Institute for Radio Astronomy (ASTRON), Postbus 2, 7990
AA Dwingeloo, the Netherlands\\
$^4$ACGC, Dept of Astronomy, University of Cape Town, Private Bag X3, 
Rondebosch 7700, South Africa} 

\date{MNRAS accepted (arXiv version 3), 21 Dec, 2012}

\begin{document}

\maketitle

\begin{abstract}
  We derive the projected surface mass distribution $\Sigma_M$ for
  spherically symmetric mass distributions having an arbitrary rotation
  curve.  For a galaxy with a flat rotation curve and an ISM disk having
  a constant Toomre stability parameter, $Q$, the ISM surface mass
  density $\Sigma_g$ as well as $\Sigma_M$ both fall off as $R^{-1}$.
  We use published data on a sample of 20 well studied galaxies to show
  that ISM disks do maintain a constant $Q$ over radii usually
  encompassing more than 50\%\ of the \HI\ mass.  The power law slope in
  $\Sigma_g$ covers a range of exponents and is well correlated with the
  slope in the epicyclic frequency.  This implies that the ISM disk is
  responding to the potential, and hence that secular evolution is
  important for setting the structure of ISM disks. We show that the gas
  to total mass ratio should be anti-correlated with the maximum
  rotational velocity, and that the sample falls on the expected
  relationship.  A very steep fall off in $\Sigma_g$ is required at the
  outermost radii to keep the mass and angular momentum content finite
  for typical rotation curve shapes, and is observed.  The observation
  that \HI\ traces dark matter over a significant range of radii in
  galaxies is thus due to the disks stabilising themselves in a normal
  dark matter dominated potential.  This explanation is consistent with
  the cold dark matter paradigm.
\end{abstract}

\section{Introduction\label{s:intro}}\label{s:intro}

\HI\ has long been the best tracer of Dark Matter (DM) in galaxies
\citep[e.g.][]{bosma81,vanderhulst+93}. This is because it typically
extends much further than the optically bright portion of a galaxy, in a
fairly regular disk.  But this ability to trace DM seems uncanny:
typically the projected DM surface density scales very well with the
measured \HI\ surface density
\citep{bosma81,sancisi83,cb89,ccbv90,cp90a,cp90b,jc90,broeils92,mcbf96,hvs01}. \citet{mestel63}
had already shown that the flat rotation curve typically seen in the
outer parts of spiral galaxies requires a surface mass density
$\Sigma(R) \propto\ R^{-1}$ fall-off in the disk, if that is where the
dominant mass is located.  This is not what is observed in the optical
but close to the behaviour of \HI\ disks. \HI\ not only is a good
dynamical tracer of DM but its distribution scales linearly with the DM
in the outskirts of galaxies.  This has prompted some researchers to
posit that DM may be gaseous, perhaps in a disk configuration
\citep{pcm94,pc94,gs96,pr05,bournaud+07,hz11}.  The scaling between DM
and \HI\ is also well explained by the Modified Newtonian Dynamics
(MOND) hypothesis in which the gravitational force law is modified; in a
MOND analysis, \HI\ is the only significant mass in the outskirts of
galaxies \citep[e.g.][]{sm02}.  Either explanation poses a problem for
the standard Cold Dark Matter (CDM) scenario.  A gaseous form of
baryonic DM would be dissipative, whereas in the CDM scenario DM only
interacts via gravity and so is non-dissipative.  The MOND scenario
requires no DM.

We propose an explanation for the linear scaling between \HI\ and
DM that is consistent with the CDM scenario.  
The ISM distribution in disks is configured to maintain a uniform
minimum stability over as much of the disk as possible.  For a flat
rotation curve, this will result in a surface density profile having the
same form as the dominant mass.  In \S~\ref{s:q} we give the basis of
our model. In \S~\ref{s:qmeas} we gather recent \HI\ rotation curve (RC)
data to test this hypothesis.  \S~\ref{s:disc} discusses our results and
presents our conclusions.

\section{The structure of dynamically stable gas dominated disks}\label{s:q}

The stability of a purely stellar or purely gaseous disk is a well
studied problem starting with the work of \citet{safronov60} and
\citet{toomre64}.  For a disk to be stable
against axi-symmetric perturbations, the support from a combination of random motions and
centrifugal forces must be larger than the gravitational
attraction. This is expressed as a ratio, the ``Toomre $Q$'' parameter:
\begin{equation}
Q \equiv \frac{\sigma \kappa}{\pi G \Sigma}\label{e:q}
\end{equation}
where $\Sigma$ is the mass density in the disk, $\sigma$ is the velocity
dispersion, and $\kappa$ is the epicyclic frequency given by
\begin{equation}
\kappa = \frac{V}{R}\sqrt{2\left(1 + \frac{R}{V}\frac{dV}{dR}\right)}\label{e:kappa}
\end{equation}
with $V$ being the rotational velocity at radius $R$. For a purely
gaseous disk to be stable $Q > 1$, while a purely stellar disk requires
$Q > 1.07 = 3.36/\pi$ for stability.  Unstable disks result in the
formation of bars and spiral arms which gather the ISM, enhancing star
formation efficiency and thus reheating the disk through feedback
\citep{hohl71,sc84,debattista+06}.  The stability of a multi-component
disk, i.e. gas and stars, is more complex \citep{js84,rafikov01,rw11}.
Here we are concerned primarily with the outer disk, where the ISM mass
dominates.  In those cases the single fluid $Q$ given by eq.~\ref{e:q} is
sufficient for our purposes.  In some of the galaxies we analyse the
stellar disk {\em is\/} important, and we show that the results are
usually not significantly different over the radii we are concerned with
when a multi-component disk analysis is employed.

As done commonly, we consider a galaxy where the dominant mass has a
spherical distribution of mass with density $\rho_M(R)$, which also
contains an embedded ``light'' (low mass) gas disk having surface mass
density $\Sigma_g$. We take the projected {\em total\/} surface mass
density $\Sigma_M(R)$ to be
\begin{equation}
\Sigma_M(R) = 2R\rho_M(R).\label{e:M}
\end{equation}
This is the projection of a spherical shell onto a ring of the same
radius at the equator.  This is the correct definition when we
are concerned with the rotation curve (hereafter RC, or $V$) of the
dominant mass.  Under our assumptions, disk
particles in circular orbits only feel the potential of the mass
interior to their orbit.  We assume this geometry precisely
because it allows some simple derivations.  \citet{om00} note
that theory indicates that DM halos should be somewhat flattened
spheroids, but that the limited observations do not clearly state what
the flattening typically is.  If the DM is in a disk, as
might be expected for gaseous form of DM, then the spherical
approximation does not hold.  As can be discerned from Fig.~1 of
\citet{hz11}, assuming a spherical geometry will cause us to
underestimate $\Sigma_M(R)$ at large $R$ in this case.  However, for typically assumed RC
shapes and beyond a few core radii, the ratio of true to inferred
$\Sigma_M(R)$ changes slowly and only by $\sim 10$\%.  Hence the
simplifying assumption of a spherical halo is not critical to our
results.

Assuming standard Newtonian gravity and pure circular orbits, it is
straight-forward to derive $\Sigma_M(R)$ as a function of the RC
\begin{equation}
\Sigma_M = \frac{1}{2\pi G R}\frac{d(RV^2)}{dR}
\end{equation}
which becomes
\begin{equation}
\Sigma_M = \frac{1}{2\pi G}\frac{V}{R}\left(2R\frac{dV}{dR} + V\right).\label{e:SM}
\end{equation}
We define $\zeta$ to be the local gas to total mass ratio
\begin{equation}
\zeta \equiv \frac{\Sigma_g}{\Sigma_M}.
\end{equation}
The case of $\zeta$ being constant corresponds to the gas disc tracing
the total mass.  We will also have occasion to consider the integrated
quantity
\begin{equation}
{\cal Z} = \frac{M_g}{M(R_{\rm max})}
\end{equation}
where $M(R_{\rm max})$ is the mass within the maximum radius, $R_{\rm
  max}$, of the \HI\ distribution.

We emphasise that the mass densities $\Sigma_M$, and $\rho_M$ in
eq.~\ref{e:M} and throughout are the projected and spherically averaged
mass densities of {\em all\/} mass, i.e.\ disk (stars and gas), bulge,
and halo (luminous and dark). Likewise $\zeta$ and ${\cal Z}$ are the
local and integrated gas to {\em total\/} mass ratio.  We adopt these
definitions to simplify the analysis, avoiding the need to fit the
rotation curve into contributions from the different components.  The
quantity $\zeta$ is thus the reciprocal of the ratio plotted in Fig.~7
of \citet{bosma81} where it first became apparent that \HI\ traces
whatever dominates the mass in the outskirts of galaxies.  The
implications of our results stem from the well established result that
at large radius the dominant form of mass in galaxies is DM
\citep[e.g.][]{freeman70,rubin+78,fg79,kent87,cb89,ccbv90,deblok+08}.

Solving eq.~\ref{e:q} for
$\Sigma_g$ and employing eq.~\ref{e:kappa} and eq.~\ref{e:SM} yields
\begin{equation}
\zeta = \frac{2\sigma}{Q}
         \frac{\sqrt{2[1+(R/V)(dV/dR)]}}{V+2R(dV/dR)}.\label{e:zetadef}
\end{equation}
It is instructive to adopt a power law form to the RC
\begin{equation}
V \approx k R^\gamma,
\end{equation}
where $k$ is a constant.  This approximation works fairly well over a
limited range of radii in galaxies, which is sufficient for our
purposes.  This results in 
\begin{equation}
\zeta = \frac{\sigma}{Q}\frac{\sqrt{8(1+\gamma)}}{kR^\gamma(2\gamma+1)}.\label{e:zg}
\end{equation}
The power law index $\gamma$ has a narrow range of allowed values:
\begin{equation}
-\frac{1}{2} \leq \gamma \leq 1.
\end{equation}
A $\gamma$ below this range means the RC drops faster than Keplerian,
while $\gamma$ above this range requires a ``hollow'' mass distribution.  
$\gamma = -1/2$ corresponds to $\zeta = \infty$ since this requires $\Sigma_M
= 0$.  In the outer disks of many disk galaxies the RC is flat at its
maximum value, hence $\gamma = 0$, $V = V_{\rm max}$, and thus
\begin{equation}
\zeta = \frac{\sqrt{8}}{V_{\rm max}}\frac{\sigma}{Q}.\label{e:zflat}
\end{equation}
If $\sigma$ and $Q$ are also constant, or their
ratio is, then $\zeta$ is also constant and the gas disk will track the
total mass distribution.  From eq.~\ref{e:SM} then
\begin{equation}
\Sigma_M(R) = \frac{1}{2\pi G}\frac{V_{\rm max}^2}{R}, \\
\Sigma_g(R) = \frac{\sqrt{2}\sigma}{\pi Q G}\frac{V_{\rm max}}{R}.\label{e:sgbigr}
\end{equation}
This is the well known relationship of the total surface mass density 
falling off as $R^{-1}$ where the RC is flat, which is a
fair approximation of what is observed in most spiral galaxies. 

We posit that disks evolve towards maintaining a constant $Q$.  Simple
feedback should encourage such a condition. Over the optical face of a
galaxy, star formation is likely to be the regulating agent.  Regions of
the disk where $Q$ is higher than average have a disk that is a
combination of hot or under-dense compared to their surroundings.  In
these regions any star formation activity would decrease (relative to
their surroundings), lowering $\sigma$, thus allowing more ISM to
accumulate or cool and decreasing $Q$.  In regions where $Q$ is low, the
disk is a combination of cold or over-dense.  Star formation will be
enhanced in these areas, increasing $\sigma$ as feedback from the newly
born stars kicks in.  While the outer disk is usually considered to be
devoid of star formation, the discovery of outlying \HII\ regions
\citep{fwgh98} and XUV disks
\citep{thilker+05,thilker+07} indicates that in many cases there are
sources of new stars that can help regulate disks.  Even in the case of
pure gaseous disks, the simulations by \citet{wmn02} show that a turbulent clumpy
disk develops with a large range of $Q$ over short scales, but with a
quasi-steady equilibrium maintained with little variation in average $Q$
with time or radius.

\section{Outer disk Q measurements}\label{s:qmeas}

We gathered data on 20 galaxies to test our hypothesis that outer disks
maintain a nearly constant $Q$.  The majority of the data is from {\em
  The \HI\ Nearby Galaxy Survey\/} (THINGS) for which detailed published
RCs can be found in \citet{oh+08} and \citet{deblok+08}.  Data from
individual studies of six additional galaxies with extended disks are
included to test the robustness of the results
\citep{csr09,edk10,gentile+07,sorme10,wjk04,wbk10}.  The main criteria
for selection is that the galaxy have tabulated data available from
recent studies (within $\sim 5$ years) and \HI\ profiles
extending beyond the optical radius $R_{25}$ (where the $B$ band surface
brightness is 25 mag arcsec$^{-2}$).  Table~\ref{t:samp} lists basic
quantities of the sample and the data sources, arranged
by the maximum rotational velocity $V_{\rm max}$.  This
sample covers $50\, {\rm km\,s^{-1}} \lapeq V_{\rm max} \lapeq 375\,
{\rm km\,s^{-1}}$ and the full range of late-type galaxy morphologies
including spirals from types S0 (NGC~1167) to Sd (e.g. NGC~300),
irregulars of types Sm (IC~2574) and Im (e.g.\ NGC3741) as well as Blue
Compact Dwarf (BCD) galaxies (e.g.\ NGC~2915).  The majority of the
sample is nearby, with only two having distance $D >
15$ Mpc.  Therefore, we adopt $D$ values that are not based on the
Hubble flow, where possible.  The sources of $D$ 
are given in Table~\ref{t:samp}.  For the two cases where we use $D$
based on redshift, we adopt the model given by NED\footnote{The
  NASA/IPAC Extragalactic Database (NED) is operated by the Jet
  Propulsion Laboratory, California Institute of Technology, under
  contract with the National Aeronautics and Space Administration.} for
the Hubble flow corrected for inflow to the Virgo cluster, Great
Attractor, and Shapley super-cluster \citep{mould+00} and standard
cosmological parameters $H_0 = 73\, {\rm km\, s^{-1}\, Mpc^{-1}}$,
$\Omega_{\rm matter} = 0.27$, and $\Omega_{\rm vacuum} = 0.73$

\begin{table*}
  \caption{Sample properties}\label{t:samp}
  \begin{tabular}{@{}lrrrrrrl}
    \hline
    Galaxy & morphology & $D$ & $R_{25}$ & $R_{\rm max}$ & \MHI & $V_{\rm max}$ & Data ; distance source \\
    (1) & (2) & (3) & (4) & (5) & (6) & (7) & (8) \\
    \hline
    DDO154    & IB(s)m     &  4.3 &  1.24 &  8.28 &  0.45 &  50 & \citet{deblok+08,makarova+98} \\
    NGC3741   & Im         &  3.2 &  0.97 &  6.97 &  0.22 &  52 & \citet{gentile+07,dalcanton+09} \\
    ESO215    & Im$^a$     &  5.2 &  0.80 & 10.82 &  1.46 &  52 & \citet{wjk04,karachentsev+07} \\
    NGC2366   & IB(s)m     &  3.4 &  2.20 &  8.20 &  0.81 &  58 & \citet{oh+08,dalcanton+09} \\
    IC2574    & SAB(s)m    &  3.8 &  7.26 & 11.07 &  1.67 &  78 & \citet{oh+08,dalcanton+09} \\
    NGC2915   & BCD$^b$    &  4.1 &  1.23 & 10.12 &  0.49 &  86 & \citet{edk10,meurer+03} \\
    NGC300    & SA(s)d     &  2.0 &  5.81 & 19.36 &  2.55 &  99 & \citet{wbk10,freedman+01} \\
    ADBSJ1138 & BCD$^c$    & 50.0 &  2.27 & 24.20 &  1.92 & 106 & \citet{csr09}; NED \\
    NGC7793   & SA(s)d     &  3.9 &  5.78 &  7.74 &  1.11 & 118 & \citet{deblok+08,karachentsev+03b} \\
    NGC2403   & SAB(s)cd   &  3.2 &  7.52 & 18.01 &  3.59 & 144 & \citet{deblok+08,freedman+01} \\
    NGC3198   & SB(rs)c    & 13.8 & 13.22 & 37.51 & 14.50 & 159 & \citet{deblok+08,freedman+01} \\
    NGC3621   & SA(s)d     &  6.6 &  9.57 & 25.77 &  9.43 & 159 & \citet{deblok+08,freedman+01} \\
    NGC4736   & (R)SA(r)ab &  4.7 &  5.41 &  9.61 &  0.52 & 198 & \citet{deblok+08,karachentsev+03} \\
    NGC6946   & SAB(rs)cd  &  6.8 & 10.79 & 22.08 &  7.00 & 224 & \citet{deblok+08,ksh00} \\
    NGC5055   & SA(rs)bc   &  7.9 & 14.30 & 38.15 &  8.29 & 212 & \citet{deblok+08,tully+09} \\
    NGC2903   & SA(rs)bc   &  8.9 & 15.50 & 29.34 &  6.45 & 215 & \citet{deblok+08,dk00} \\
    NGC3521   & SAB(rs)bc  & 10.7 & 11.49 & 31.17 & 12.70 & 233 & \citet{deblok+08} \\
    NGC3031   & SA(s)ab    &  3.6 & 11.42 & 14.80 &  4.05 & 260 & \citet{deblok+08,freedman+01} \\
    NGC2841   & SA(r)b:    & 14.1 & 28.90 & 51.68 & 13.90 & 324 & \citet{deblok+08,saha+06} \\
    NGC1167   & SA0$^-$    & 66.0 & 23.49 & 63.88 & 12.20 & 377 & \citet{sorme10}; NED \\
   \hline
  \end{tabular}
Columns: (1) galaxy name; (2) galaxy morphology from the RC3 catalogue
\citep{rc3} except as following: a.\ \citet{wjk04}, b.\ \citet{mmc94}, c.\ \citet{csr09}; (3) Distance
in Mpc; (4) radius at the $B$ band 25 mag arcsec$^{-2}$; (5) radius of
the outermost \HI\ measurement; (6) \HI\ mass in $10^9$ \Msun; (7)
Maximum rotation amplitude in km s$^{-1}$. 
\end{table*}

\begin{figure*}
\includegraphics[width=140mm]{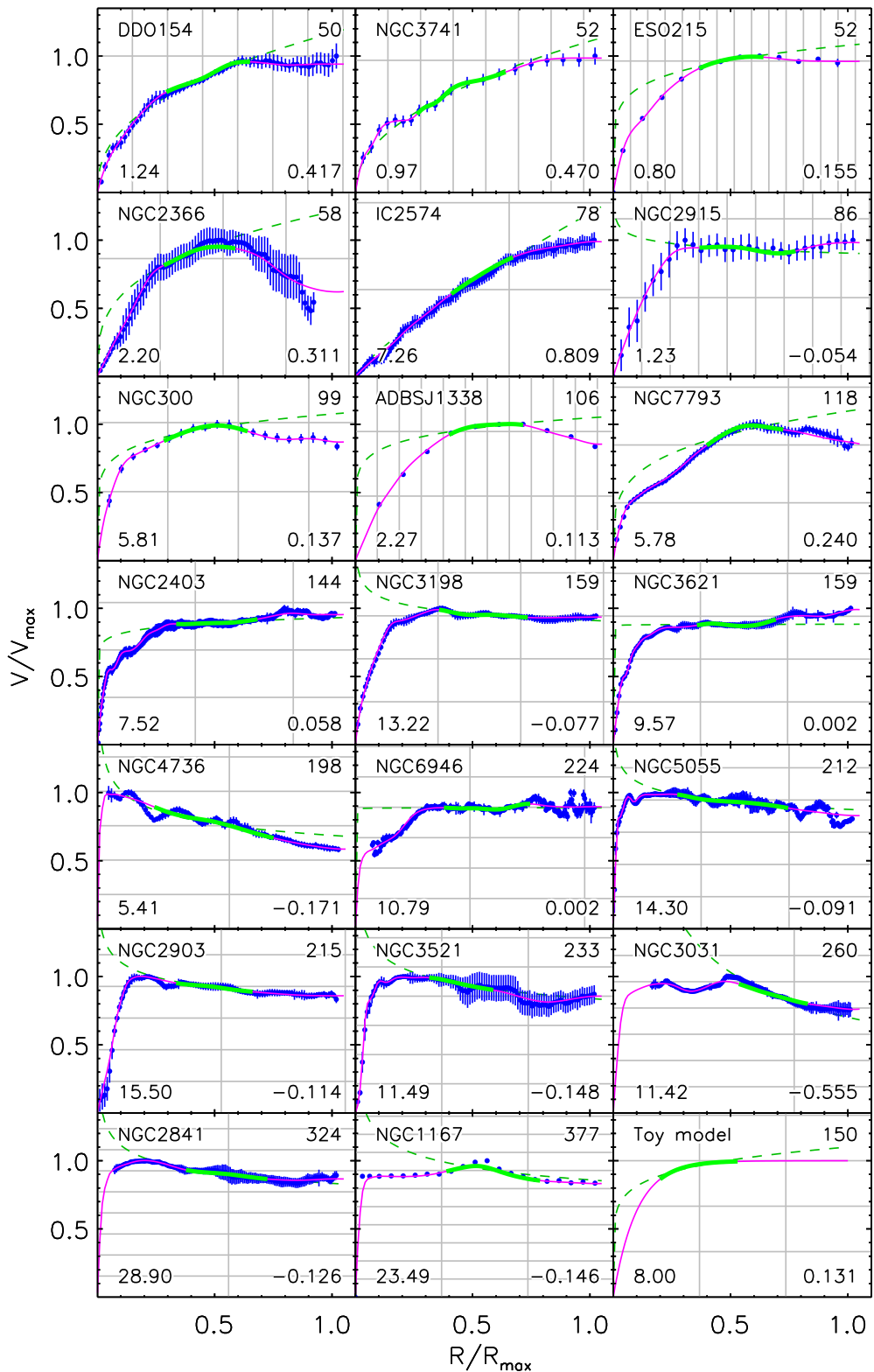}
\caption{Rotation curves (RC), ordered and scaled by
  maximum rotational velocity, $V_{\rm max}$, plotted against radius $R$
  scaled by the maximum radius of the data $R_{\rm max}$. We show the
  measured points and their errors (blue), the adopted spline RC
  fits (magenta), with the middle 50\% of the HI mass highlighted
  (green), and a power law fit to the data in this region (dashed green). 
  Each panel lists the galaxy name is at upper left, the
  optical radius $R_{25}$ in kpc at lower left, $V_{\rm max}$ in km
  s$^{-1}$ at upper right and the power law slope $\gamma$ at lower
  right. Vertical (grey) lines mark intervals of $R_{25}$, while horizontal
  lines mark intervals of 50 km s$^{-1}$. The toy model plotted in the lower
  right panel is an RC of the form used by \citet{leroy+08} with 
  $R_{\rm flat} = 4$ kpc.
  \label{f:rc}}
\end{figure*}

The main observational quantities of importance are the RC and the
$\Sigma_g$ profile.  The RCs are shown in
Fig.~\ref{f:rc}.  We fit a cubic-spline to the RCs using knots set by
eye.  In performing the fit, we keep the number
of spline knots to a minimum and try to follow the data to within the
errors. However, we smooth over small scale fluctuations in the RCs,
presumably due to spiral arms or non circular motions.  The fitted splines
are shown as continuos lines in Fig.~\ref{f:rc}. We
chose this functional form for its flexibility and because it allows for
easy evaluation of the derivative $dV/dR$ needed for the calculation of
$\kappa$ and thus $Q$.  For comparison, we plot a model RC using the
functional form adopted by \citet{leroy+08}, with parameters $V_{\rm
  flat} = 150\, {\rm km\, s^{-1}}$, $R_{\rm flat} = 0.5 R_{25}$, and
$R_{25} = 8$ kpc.  The sample includes many galaxies with flat
RCs at large radii, like this model (DDO154, 
ESO215\footnote{full name: ESO215-G?09 \citep{wjk04}.}, NGC~2915,
NGC~2403, NGC~3198, NGC~6946, NGC~2841), or are still rising (IC~2574,
NGC~3741, NGC~3621).  However, about half the sample have RCs that have
a substantial range of radii where they are declining (NGC~2366,
NGC~300, 
ADBS~J1138\footnote{full name: ADBSJ~113845+2008 \citep{csr09}.}, 
NGC~7793, NGC~4736, NGC~5055, NGC~2903,
NGC~3521, NGC~3031, NGC~1167).  To illustrate the range of shapes, we
fit a power-law RC to the data between a limited range of radii, defined
below.  These and the other power-law fits in this paper were performed
as linear fits in log-log space using the IDL procedure {\tt MPFIT}
\citep{markwardt09}.  The fits are shown as dashed lines in
Fig.~\ref{f:rc}, with the fit parameters, including zeropoint, slope
$\gamma$, and dispersion of the residuals $\epsilon$ listed in
Table~2.

\begin{figure*}
\includegraphics[width=140mm]{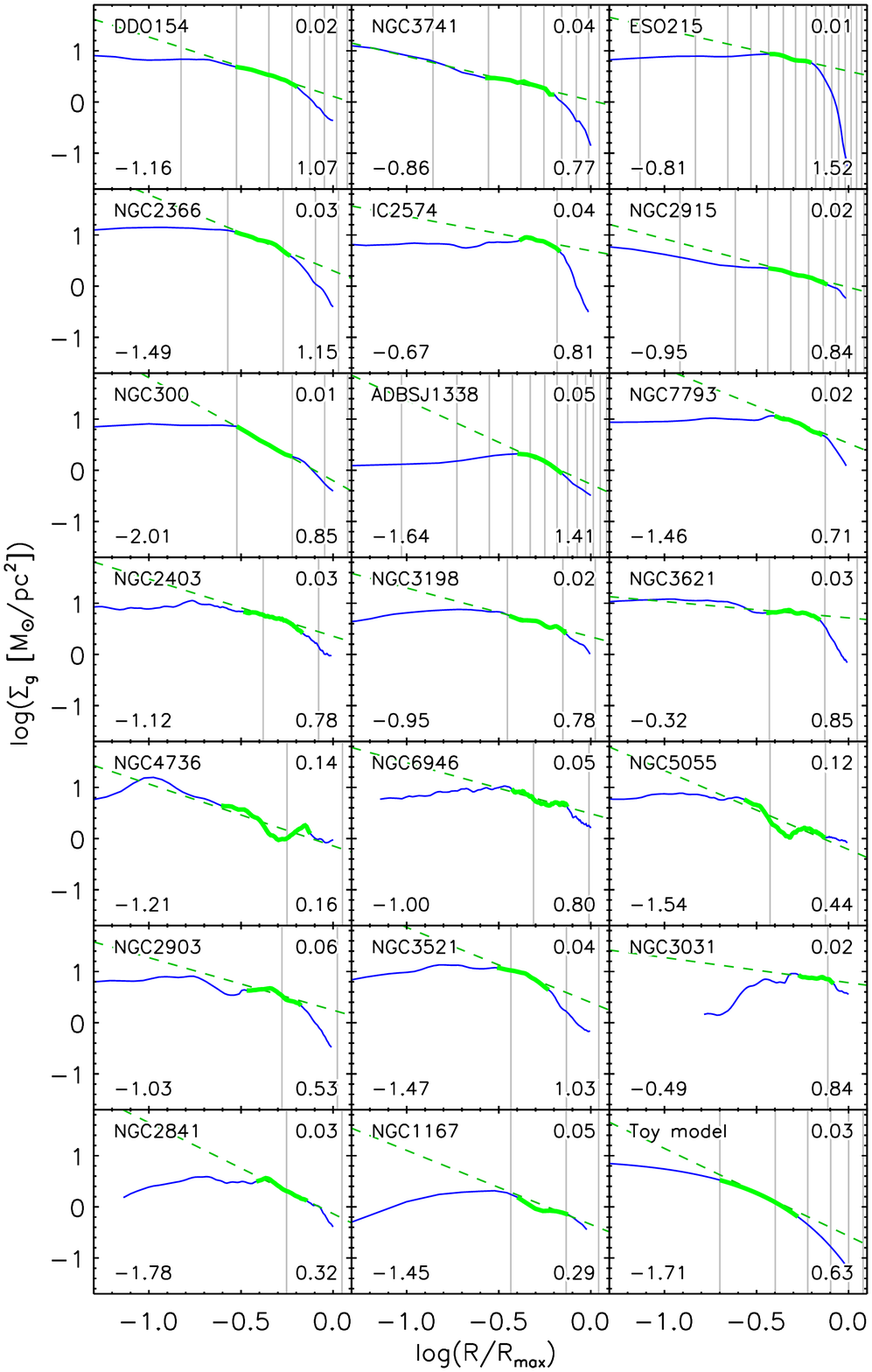}
\caption{Radial profiles of ISM surface mass density $\Sigma_g$
  (blue) plotted on a log-log scale. Line colour and highlighting are the
  same as in Fig.~\ref{f:rc}. Power law fits to the highlighted portion
  of the profile are shown with dashed lines (dark green).  Each panel
  lists dispersion of
  the residuals of the fits at upper right, the power law index $N$ at lower
  left, and $\log(\Sigma_g)$ at $R_{\rm 25}$ at lower right. The toy
  model shown in the lower right panel is an exponential profile
  (eq.~\ref{e:edisk}) with scale length $\alpha^{-1} = R_{25}$ and
  central density $\Sigma_{g,0} = 10$ \Msun\ pc$^{-2}$.\label{f:sg}}
\end{figure*}
 
The $\Sigma_g$ profiles are shown in Fig.~\ref{f:sg}.  They were derived
from the inclination corrected \HI\ profiles assuming $\Sigma_g =
1.3\Sigma_{\rm HI}$ to account for elements heavier than hydrogen.  Since we
are primarily interested in the outer disk we assume that the molecular
content is negligible.  The profiles are drawn in log-log space, to
highlight any power-law behaviour.  In general, the profiles are nearly
flat or decreasing towards small radii, and steeply dropping at the
largest radii, leaving a ``knee'' where the profile has an approximately
power law form.  We examined each profile and determined the inner and
outer radii at
which the power-law portion begins and ends, respectively, and the
fraction of the total \HI\ mass \MHI\ within those radii.  On average
21($\pm 10$)\%\ of \MHI\ is interior to the inner radius and 75($\pm
10$)\%\ is interior to the outer, hence the power law knee 
contains a bit over 50\%\ of \MHI\ on average.  In order to treat
the galaxies consistently, we determine the radii enclosing 25\%\ and
75\%\ of \MHI, $R_1$ and $R_2$ respectively, and fit a power law
\begin{equation}
\Sigma_g \propto R^N
\end{equation}
to the points between these radii.  The fits are
shown in Fig.~\ref{f:sg} with the parameters listed in the corners of
the panels and in Table~2.  The range
$R_1$ to $R_2$ is highlighted in our figures by using thicker lines to
plot the profiles.  $N$ spans the range $-2.01$
(NGC~300) to $-0.32$ (NGC~3621).  The power law fit is reasonable.  The
average $\epsilon_{\Sigma_g} = 0.042$ dex (10\%), and in only two cases
is $\epsilon_{\Sigma_g} > 0.1$ dex (in NGC~4736, NGC~5055).

While we use a power law approximation of $\Sigma_g$ for convenience,
\HI\ profiles are not always characterised as such in the literature.
\citet{fc01}, for example, use an exponential profile:
\begin{equation}
\Sigma_g = \Sigma_{g,0} e^{-\alpha R} \label{e:edisk}
\end{equation}
where $\Sigma_{g,0}$ is the central ISM surface mass density and
$\alpha^{-1}$ is the disk scale length. They show that the ISM in
their viscous disk models are well characterised by such a
profile over a wide range in radii, with the ISM scale length greater
than that of the stellar disk \citep[which is well
known to follow an exponential profile;][]{freeman70}.  The bottom
right panel of Fig.~\ref{f:sg} shows an exponential profile having
$\Sigma_{g,0} = 10\, {\cal M}_\odot\,{\rm pc^{-2}}$ and $\alpha^{-1} =
R_{25}$. This profile is similar in shape to that of our sample.
We fit a power law to this profile in the same manner as the sample
galaxies, and report the results in Table~2 and
Fig~\ref{f:sg}.  Over the relevant $R_1$ to $R_2$ a power
law fit to an exponential profile gives a $\epsilon_{\Sigma_g}$ of $0.03$ dex
(7\%). In comparison, 55\%\ of our sample have $\epsilon_{\Sigma_g}$
less than or equal to this.  Thus a power law is at least as good a
functional form as an exponential for about half the sample.

\begin{figure*}
\includegraphics[width=140mm]{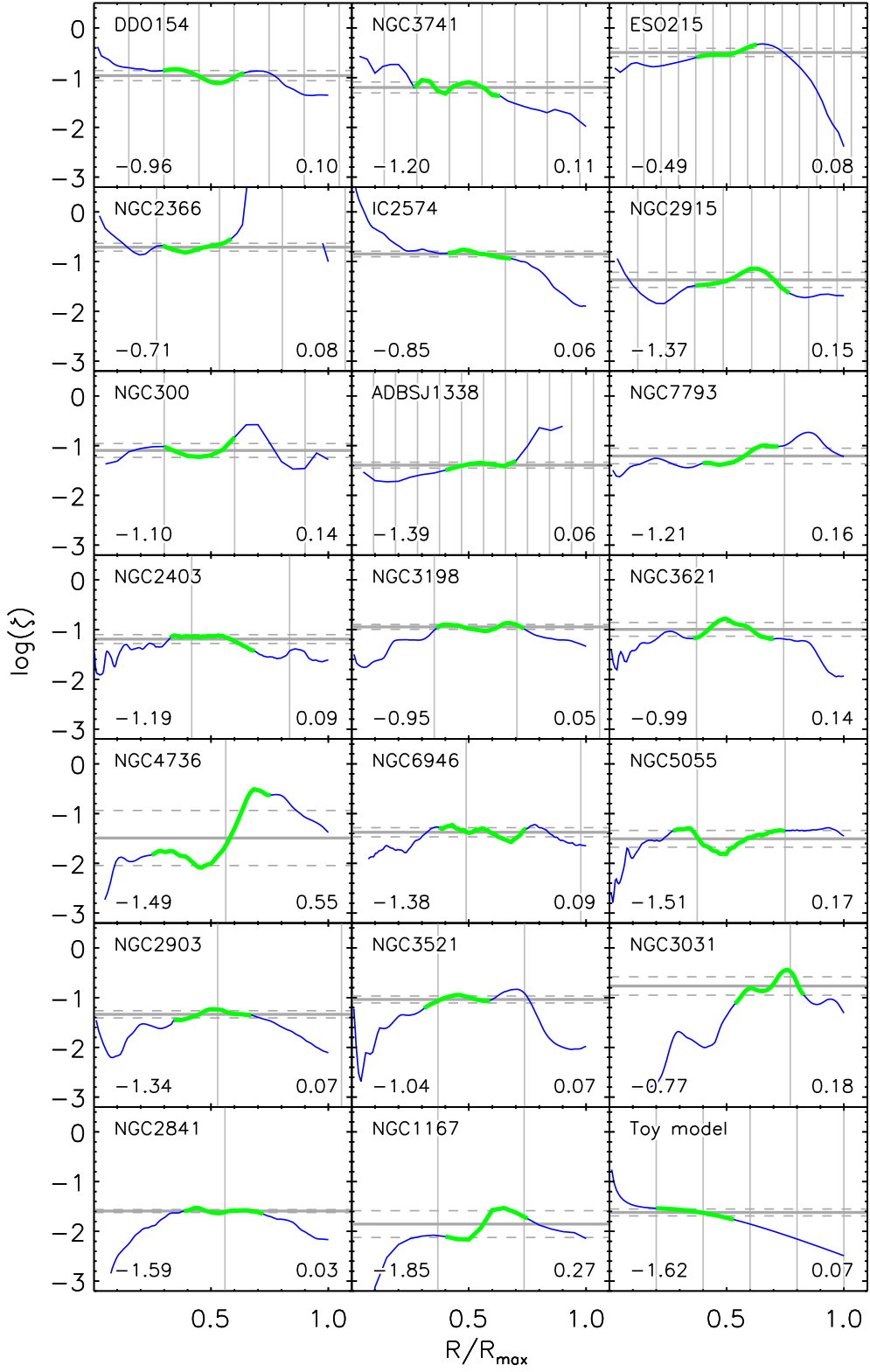}
\caption{Profiles of the ratio $\zeta$ of gas surface mass
  density $\Sigma_g$ to projected total surface mass density $\Sigma_M$,
  plotted on a log scale, against linear scaled radius. Line colour and
  highlighting are the same as in Fig.~\ref{f:rc}. The average
  $\log(\zeta)$ and dispersion about the average in the highlighted
  region is marked with the solid and dashed horizontal (gray) lines and
  listed at the lower left and lower right corners. The
  vertical lines show intervals of $R_{25}$. \label{f:zeta}}
\end{figure*}
 
Figure~\ref{f:zeta} shows the $\zeta$ profiles in order to
test the premise that \HI\ traces the total mass.  To form the ratio we
take $\Sigma_g$ from the observed profile and $\Sigma_M$ from
eq.~\ref{e:SM}.  In general the profiles are highly structured.  This is
due to irregularities in both the $\Sigma_g$ and $V$ profiles. The
$\zeta$ profiles are particularly sensitive to the latter since it
depends on the derivative of the RC (eq.~\ref{e:SM}).  The toy model,
which combines the smooth RC shown in Fig.~\ref{f:rc} and the smooth
$\Sigma_g$ profile shown in Fig.~\ref{f:sg}, has $\epsilon_\zeta =
0.07$ dex. Six of the galaxies in the sample have $\zeta$ profiles that
have $\epsilon_\zeta$ at this level or smaller.

\begin{figure*}
\includegraphics[width=140mm]{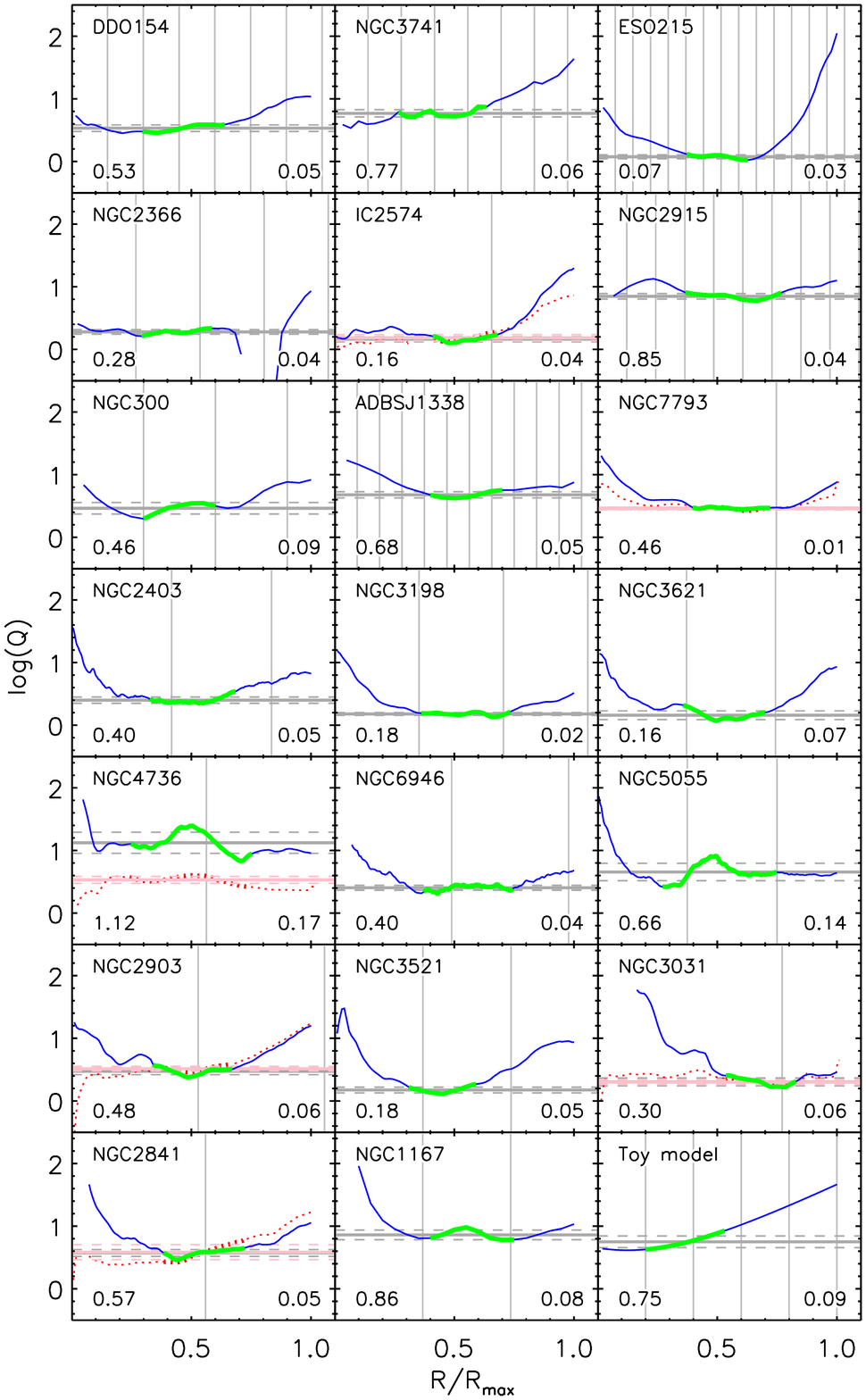}
\caption{Toomre stability parameter $Q$ profiles on a log scale,
  plotted against linear scaled radius. Line colour and highlighting are
  the same as in Fig.~\ref{f:rc}.  The average $\log(Q)$ in the
  highlighted region is marked at lower left and the rms about this
  average is given at lower right, and these are shown with the
  horizontal (grey) lines. The vertical lines show intervals of
  $R_{25}$. In the six cases where the fit range is largely within
  $R_{25}$ we show profiles of the two fluid stability parameter \Qsg\
  \citep{rw11} as red dotted lines (thicker over the fit range); the
  pink thick line shows the average $\log(\Qsg)$ over the fit range and
  the pink dashed lines are offset from this by the rms about this
  average.  \label{f:q}}
\end{figure*}
 
Figure~\ref{f:q} shows the $Q$ profiles, for an assumed
constant velocity dispersion $\sigma = 8\, {\rm km\, s^{-1}}$.  The
average $\log(Q)$ between $R_1$ and $R_2$ and the $\epsilon$ in this
value are listed in Table~2 and the panels of
Fig.~\ref{f:q}.  For the majority of the sample $Q$ is quite flat
between $R_1$ and $R_2$, and somewhat beyond for many cases. Over this
range, the average $\epsilon_{\rm Q}$ is 0.06 dex (15\%), while in two
cases it is more than twice that: NGC~4736 and NGC~5055, for which the
variation in $Q$ corresponds largely to irregular structure in the
$\Sigma_g$ profile (Fig.~\ref{f:sg}). For NGC~2366, $Q$
becomes undefined for much of the disk beyond $R_2$ because, in this
case, its RC is declining more steeply than a Keplerian decline.  Since this is
unphysical, it is likely that a warp or non-circular motions cause the
RC of \citet{oh+08} to be underestimated.  In the bottom right panel we
combine the toy models shown in Fig.~\ref{f:rc} and Fig.~\ref{f:sg}.  This
model has $\epsilon_{\rm Q} = 0.09$ dex, while 70\%\ of our sample have
values less than or equal to this. As expected, $Q$ is rising beyond
$R_2$ for the majority of the sample.

The $\sigma = 8\, {\rm km\,s^{-1}}$ we adopt is in the
middle of the pack compared to what has been adopted in other studies;
e.g. \citet{kennicutt89} adopts $\sigma = 6\, {\rm km\,s^{-1}}$, while
\citet{leroy+08} use $\sigma = 11\, {\rm km\,s^{-1}}$.  Changing
to a different constant $\sigma$ will only change $Q$ by a constant multiplicative value.  Following
\citet{tamburro+09} we also performed calculations where $\sigma$ declined
linearly with radius, using their profiles for the few overlapping cases
between our study and theirs, and otherwise setting $\sigma = 10\, {\rm
  km\,s^{-1}}$ at $R_{25}$ and falling linearly with radius to
$\sigma = 5\, {\rm km\,s^{-1}}$ at the last measured point of the radial
profiles.  The resultant $Q$ profile do not look very different from 
fig.~\ref{f:q}; in particular, the decline in $\sigma$ does not remove the
rise in $Q$ often found at large $R$.

While Fig.~\ref{f:q} shows that $Q$ is fairly constant over radii
incorporating about half of the \HI, usually it increases outside this
range.  The explanation of what is happening at large radii is fairly
simple. There, RCs typically are flat or become flatter as can be seen
in Fig~\ref{f:rc}.  As shown by eq.~\ref{e:sgbigr}, for $\zeta$ to be
constant in this limit requires $\Sigma_g \propto R^{-1}$.  Such a
profile can not be maintained to arbitrary radius because it would
require infinite \MHI.  To have a disk with finite mass and angular
momentum requires a more rapid drop-off in $\Sigma_g$, as is observed
beyond $R_2$ (Fig.~\ref{f:sg}), hence $Q$ typically rises at the largest
radii.  \citet{hvs01} point out that the sharp drop off in $\Sigma_g$ at
large $R$ does not correspond to an expected decline in RCs, thus
casting doubt on the ability of \HI\ to trace DM.  \citet{hz11} counter
that the fitting technique of \citet{hvs01} over emphasises the RC fits
at large $R$ where much of the hydrogen is ionised.  Neutral or ionised,
there is a finite ISM mass, and this will limit the ability of the ISM
to trace DM.

There are multiple causes for the rising $Q$ profiles towards small
radii.  The central parts of galaxies typically have a significant
molecular content, which we have not included here, hence $Q$ is
overestimated.  In addition, the stellar disk usually dominates the
mass distribution, so that a single component $Q$ is inadequate for
determining the true disk stability \citep{leroy+08,rw11}.  Our
assumption has been that the \HI\ largey resides in the outer disk.
However, in six of the sample galaxies (IC~2574, NGC~7793, 4736, 2903, 
3031, and 2841) the fitting range is largely interior
to $R_{25}$, hence the stellar disk is likely to play an important role
in determining the stability of the disk in these cases.

In order to determine the true disk stability for these cases, we
calculated the two fluid (stars and gas) stability parameter \Qsg.
There are various formulations of the two fluid stability parameter
\citep[e.g.][]{js84,ws94,rafikov01,rw11}.  Here, we calculate \Qsg\ from
eq.~9 of \citet{rw11} which accounts for the thickness of the stellar
and gaseous disks.  It can be rewritten in a simplified form as
\begin{equation}
\frac{1}{\Qsg} = \frac{p_\star}{Q_\star} + \frac{p_g}{Q_g}
\end{equation}
Where $Q_\star$ and $Q_g$ are the stability parameters for the stellar
and gaseous components of the disk calculated separately using
eq.~\ref{e:q}, and $p_\star$, $p_g$ are weight factors which depend on
the velocity dispersions of the stars and gas and the value of $Q_\star$
compared to $Q_g$.  Calculation of \Qsg\ requires the stellar mass
density $\Sigma_\star$ in the disk which we derive from the stellar mass
radial profiles of \citet{deblok+08}, and the radial component of the
stellar velocity dispersion ellipsoid $\sigma_{\star,r}$.  We take
$\sigma_{\star,r} \approx \sigma_{\star,z}/0.6$ following
\citet{shapiro+03} where $\sigma_{\star,z}$ is the vertical component of
the stellar velocity dispersion. This is given by
\begin{equation}
\sigma_{\star,z} = \sqrt{2\pi G \Sigma_{\rm disk} h_\star}
\end{equation}
where $\Sigma_{\rm disk} = \Sigma_\star + \Sigma_g$ is the total disk
surface mass density and $h_\star$ is the stellar disk scale height.
Following \citet{leroy+08} we estimate the scale height from the disk
scale length using $h_\star \sim l_\star/7.3$ and fit the exponential
scale length $l_\star$ over the same fit region as highlighted in our
figures.  While we are primarily are concerned with \Qsg\ over the
fitted region it is instructive to also see its behaviour beyond this
range.  In the central regions, in cases where the measured
$\Sigma_\star$ is greater than the extrapolated exponential fit, we
replace $\Sigma_\star$ with this fit, under the assumption that the
excess light represents a bulge or thick disk with a scale height larger
than the disk and thus has a lesser contribution to the disk potential
than is expected for the surface brightness.  Similarly, at large $R$,
beyond the last measured $\Sigma_\star$, we also adopt the extrapolated
fit when calculating \Qsg.  The results of the \Qsg\ calculations are
shown in Fig~\ref{f:q} as the red dotted lines.  The average
$\log(\Qsg)$ over the fitted range is shown as the thick pink lines and
the dashed pink lines are offset from this line by its rms.

Figure~\ref{f:q} demonstrates that the \Qsg\ profiles are flatter than
the $Q = Q_g$ profiles when the entire radial range is considered.  The
biggest difference is for NGC~4736 where the average $\pm$ dispersion is
$\langle\log(\Qsg)\rangle = 0.53\pm 0.06$ over the fitted region; the
dispersion is a factor of three lower than that for the gas $Q$.  For
the remaining cases, the difference between $Q$ and are \Qsg\ more
subtle, over the fitted range where we find $\langle \log(\Qsg)\rangle =
0.17 \pm 0.05$, $0.46 \pm 0.02$, $0.052 \pm 0.04$, $0.30 \pm 0.04$, and
$0.59 \pm 0.12$ for IC~2574, NGC~7793, 2903, 3031, and 2841
respectively.  The dispersion in $\log(\Qsg)$ is smaller than that of
$\log(Q)$ for half the cases: NGC~4736, 2903, and 3031.  In most
cases $Q$ is an adequate proxy for the total disk stability parameter
\Qsg\ at the radii we are interested in here, even when much of the \HI\
is within the optical radius.  However, a more sophisticated analysis,
such as using \Qsg\ is required to extend the analysis even closer to
the center, or in cases like NGC~4736 where the stellar disk strongly
dominates the gaseous disk at all radii.  In \citet{zheng+12} we
perform such an analysis over the optically bright portion of galaxies,
using various prescriptions for \Qsg.  There we show that the assumption
of a constant stability disk can explain the relative distributions of
gas, stars and star formation over much of the optically bright portion
of galaxies.

\begin{figure*}
\includegraphics[width=150mm]{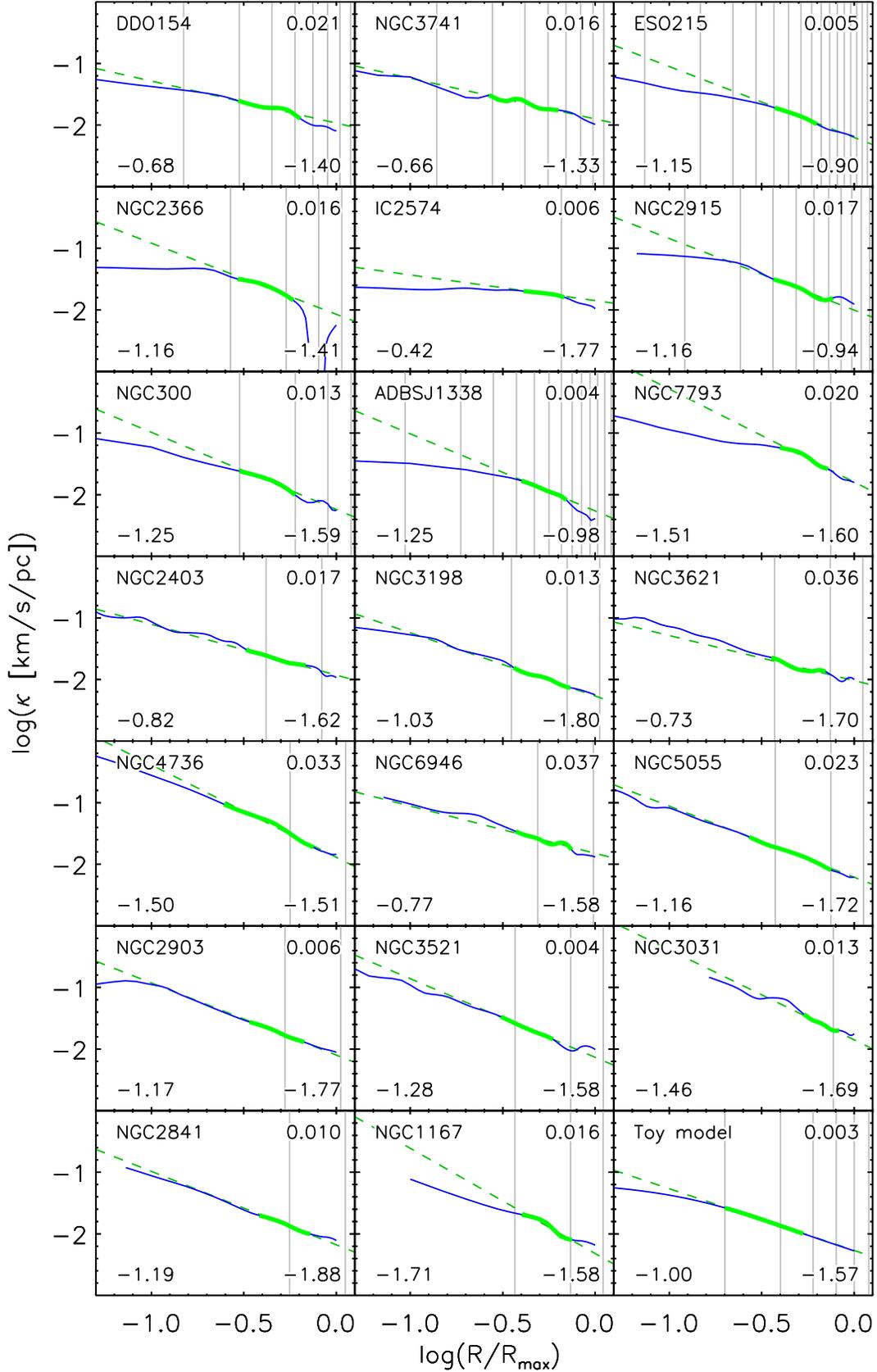}
\caption{Epicyclic frequency $\kappa$ profiles on a log-log scale. Line
  colour and highlighting are the same as in Fig.~\ref{f:rc}. 
  In each panel, the rms of the power-law fits (green
  dashed lines) are listed at upper right, the power law index $M$ at
  lower left, and $\log(\kappa)$ at $R_{\rm 25}$ at lower
  right.\label{f:k}}
\end{figure*}

Close examination of Fig.~\ref{f:rc} and Fig.~\ref{f:sg} shows that
another of our premises is not exactly correct: the RCs are not always
flat.  Instead there is a rather wide distribution of rotation curve
power law indices $\gamma$.  What then causes the nearly constant $Q$
seen in Fig.~\ref{f:q}?  This requires the numerator and denominator in
the defining equation (eq.~\ref{e:q}) to have the same shape, i.e. the
same slope.  For a constant $\sigma$, this requires the $\kappa$ and
$\Sigma_g$ profiles to have the same shape.  To test this, we fit a
power law $\kappa \propto R^M$ to the epicyclic frequency profile. The
$\kappa$ profiles and the fits to them are shown in Fig.~\ref{f:k}, with
the relevant fit parameters listed in Table~2 and the panels
of Fig.~\ref{f:k}.  A power law form is a reasonable approximation to
the $\kappa$ profiles between $R_1$ and $R_2$.

\begin{figure}
\includegraphics[width=70mm]{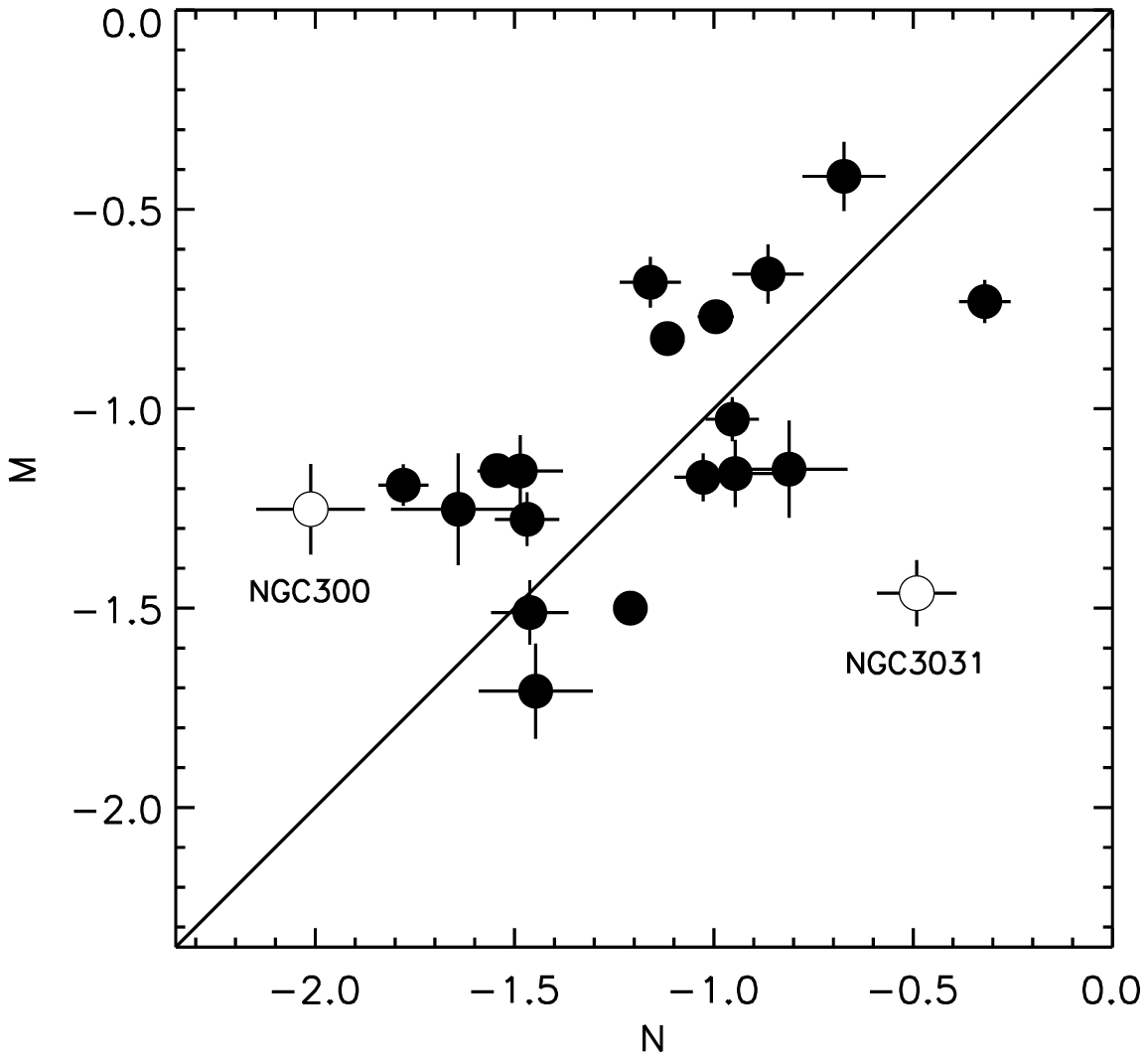}
\caption{Index $N$ of the fit to the $\Sigma_g$ profiles in
  Fig.~\ref{f:sg} plotted against the index $M$ of the fit to
  the $\kappa$ profiles shown in Fig.~\ref{f:k}.  The diagonal line
  marks $N = M$. Outliers are plotted as hollow circles, are labelled,
  and discussed further in the text.\label{f:mn}}
\end{figure}

 Figure~\ref{f:mn} plots the power law index in $\Sigma_g$, $N$ against
the power law index in $\kappa$, $M$.  There is a crude correlation
between the indices; the Pearson's correlation coefficient $r_{xy} =
0.45$, with a 2\%\ chance that the correlation is random. Examination of
the figure shows that there are two obvious outliers, and we can see
plausible reasons for their discrepant behaviour in each case.  NGC~3031
(M81) is in a nearby highly interactive group with three close
companions (M82, NGC~3077, and HoIX).  These may affect the outer ISM
distribution of NGC~3031 through either stripping material, or having
material stripped from them.  NGC~300 has a very extended and lopsided
\HI\ distribution and the steepest $N = -2.01$ in our sample.
\citet{wbk10} note that there are morphological signs of ram-pressure
stripping of the outer disk, which could steepen the $\Sigma_g$ profile.
They also note that their ATCA data may not capture the total \HI\ flux
due to missing short spacings.  The $\Sigma_g$ profile we use was
derived from a data cube that combined the ATCA data of \citet{wbk10}
and single dish \HI\ data from the Parkes 64m telescope obtained for the
GASS project \citep{kalberla+10}, thus recovering the \HI\
flux missing from the ATCA observations\footnote{Without the Parkes data
  we would have an even steeper $N = -2.36$.}.  Excluding these two
galaxies $r_{xy} = 0.63$, with a 0.3\%\ chance of being random. We
conclude that there is modest correlation between the indices which
scatter about the line $N = M$, that is the $\kappa$ and $\Sigma_g$
profiles have the same shape.  This is exactly what is needed for a
constant $Q$ disk.

\begin{figure}
\includegraphics[width=70mm]{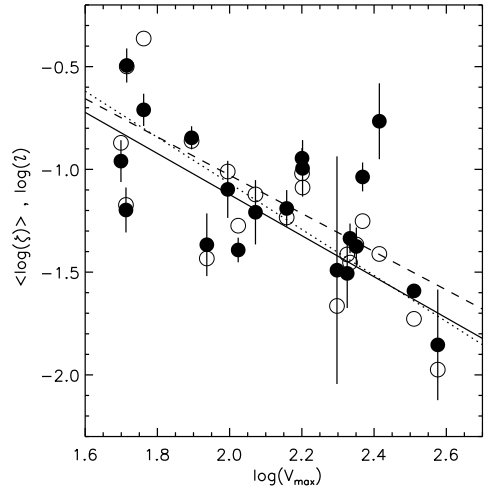}
\caption{Average $\log(\zeta)$ between $R_1$ and $R_2$,
  $\langle\log(\zeta)\rangle$, plotted as filled
  circles against rotation curve maximum $V_{\rm max}$. The
  error bars represent $\epsilon_\zeta$.  The hollow circles represent
  the integrated ratio \Zeta\ for the same galaxies.  The solid line shows the expected
  relationship for galaxies with a flat RC, eq.~\ref{e:zflat}, and
  $\sigma = 8\, {\rm km\, s^{-1}}$, $Q = 3$. The dashed line shows the
  fit to the $\langle\log(\zeta)\rangle$ data reported in
  eq.~\ref{e:zavgfit}, while the dotted line shows the fit to the
  \Zeta\ data reported in eq.~\ref{e:ztotfit}. \label{f:zv}}
\end{figure}

 A further test of the hypothesis that galaxy disks evolve to constant
$Q$ is given by eq.~\ref{e:zflat} which predicts that $\zeta$ and
$V_{\rm max}$ should be anti-correlated.  These quantities are plotted
(in the log) in Fig.~\ref{f:zv}. The filled circles show the average
$\log(\zeta)$ (i.e.\ $\langle\log(\zeta)\rangle$), between $R_1$ and
$R_2$ as marked in fig.~\ref{f:zeta}.  The expected anti-correlation is
clearly present, with $r_{xy} = -0.60$; the probability that this is due
to random sampling of uncorrelated data is less than 0.3\%.  The solid
line shows the expected anti-correlation for galaxies with a flat RC,
eq.~\ref{e:zflat}, and for the assumed $\sigma = 8\, {\rm km\, s^{-1}}$
and $Q = 3$ (this corresponds to the average $\log(Q)$ for our sample).
The dashed line is a $\chi^2$ fit to the data:
\begin{equation}
  \langle\log(\zeta)\rangle = (0.83\pm 0.13) - (0.93\pm 0.06)\log(V_{\rm max}).
  \label{e:zavgfit}
\end{equation}
The dispersion about this fit $\epsilon_{\log(\zeta)} = 0.28$ dex, while
the average offset between the expected relation and this fit is 0.08
dex, very close to what is expected from the standard error on the mean,
0.06 dex. This
demonstrates that eq.~\ref{e:zflat} provides a reasonable model for the
average gas to total mass ratio between $R_1$ and $R_2$.

The hollow circles in Fig.~\ref{f:zv} show the ratio \Zeta\ plotted
against $V_{\rm max}$.  These quantities are even more strongly
anti-correlated with $r_{xy} = -0.80$ and a less than 0.002\%\ chance
that the correlation is due to random sampling of uncorrelated data.
The dotted line shows an equally weighted least squares fit to the data
\begin{equation}
  \log({\cal Z}) = (1.17\pm 0.19) - (1.12\pm 0.09)\log(V_{\rm max});
  \label{e:ztotfit}
\end{equation}
the dispersion about this fit $\epsilon_{\log(\zeta)} = 0.24$ dex. This
is tighter than eq.~\ref{e:zavgfit}, probably because \Zeta\
is better defined than $\langle\log(\zeta)\rangle$ which is more
susceptible to noise in the $\Sigma_g$ and $V$ profiles.  

\section{Discussion and conclusions}\label{s:disc}

The correlation between the shape of the $\kappa$ and $\Sigma_g$
profiles shown in Fig.~\ref{f:mn} is profound.  Since $M$ is determined
purely by the potential, this implies that the ISM disk is responding to
the potential, and hence that secular evolution is driving the
correlation.  An alternate hypothesis is that the structure of galactic
disks is set by the mass and angular momentum accretion history
\citep{barnes02,sfov08}. However, it is not clear why there should be
any such correlation under this scenario.  CDM simulations indicate that
Milky-Way mass galaxies have had typically only two major merger event
since $z \sim 2$ \citep{dmm08,chfp08}, 10 Gyr ago, which agrees with
observations of mergers \citep{crm08}.  So, to the extent the accretion
is from major-mergers, 
the outer disks should be very lumpy.  Our sample does include some
lumpy disks (e.g. NGC~4736 and NGC~5055) and also some galaxies that may
currently be interacting (NGC~300, NGC~3031) perhaps contributing to the
scatter and outliers in Fig.~\ref{f:mn}.  The fact that the majority of
the sample falls on the $N = M$ line suggests that most are not strongly
interacting and that prior interactions happened far enough in the past
that the disk has re-stabilised and smoothed out to trace $\kappa$.  The
timescale for doing this is the orbital time $t_{\rm orb}$ ($\sim
3t_{\rm dyn}$ where $t_{\rm dyn}$ is the dynamical timescale). For our
sample, the average $t_{\rm orb}(R_2) = 0.6$ Gyr.  Taking the separation
between major merger events to be 3 Gyr, there should be approximately five
orbits between mergers, sufficient time for structure in the disk to
smooth out.  Alternatively, if accretion is slow ``cold accretion''
\citep{kkwd05,sfov08} the disk would not be expected to be lumpy.
However, that scenario does not provide an obvious explanation for the
$\Sigma_g$ and $\kappa$ profiles following each other.  Of course, cold
accretion combined with feedback to equalize $Q$, as advocated here, is
consistent with our results.

Previous studies have noted the large scatter in the HI to total or DM ratio
\citep[e.g.][]{bosma81,hvs01}. This is hard to explain in the context of
\HI\ being a linear tracer of a baryonic DM \citep{hvs01}.  We show that
an inverse correlation between $\zeta$ and $V_{\rm max}$ is expected from
constant $Q$ disks (eq.~\ref{e:zflat}), and indeed is observed
(Fig.~\ref{f:zv}).  This avoids the need to place implausibly large
quantities of molecular gas in disks.  By linking a luminous component
of the disk to the rotational velocity the \Zeta\ -- $V_{\rm max}$
relation is reminiscent of the Tully-Fisher relation \citep{tf77} and
the baryonic Tully-Fisher relation \citep{freeman99,msbd00}. However,
since \Zeta\ is a ratio, it does not provide a means to estimate ISM mass
from $V_{\rm max}$.  To do so, one needs to know the extent of the \HI\
mass distribution, since in the case of $\gamma = 0$ the local $\zeta$
and the integrated gas to total mass ratio within $R$ remain constant
and equal to each other.  Alternatively, if one has \MHI\ and $V_{\rm
  max}$ one can estimate the extent of the ISM distribution.  This is
typically the case in blind \HI\ surveys, such as HIPASS
\citep{barnes+01}, and ALFALFA \citep{giovanelli+05a} where the \HI\
flux is known but the source is unresolved.  Then one can use
eq.~\ref{e:ztotfit} to estimate \Zeta\ and from that the maximum \HI\
extent
\begin{equation}
R_{\rm max} = \frac{G \MHI}{{\cal Z} V_{\rm max}^2}.
\end{equation}
This may be useful for determining whether follow-up observations of a
particular galaxy are likely to be fruitful, or estimating the
covering factor of \HI\ absorbers \citep[e.g.][]{zvbvr05}.

Our results imply that secular processes are important for setting disk
structure.  \citet{lp74} noted that viscous disks should evolve so that
mass is concentrated in the centre and angular-momentum goes to
infinity.  As dissipative encounters cause ISM mass to be
funnelled towards the galaxy centre, the disk must also spread to conserve angular
momentum, and the ISM disk size should grow with time
\citep[e.g.][]{fc01}.  High resolution simulations of galactic disks
show that transient density waves can increase the
size of disks and alter their metallicity distributions (Ro{\v s}kar et
al.\ 2008a,b).\nocite{roskar+08a,roskar+08b} The flattening of metallicity gradients
\citep{werk+10,wpms11} in outer disks may be evidence of ISM circulation
in spreading disks.  If disks are spreading now, they should
have been denser and more compact in the past.  This would imply a
higher molecular fraction from the increased hydrostatic pressure, as
well as increased total gas content (to account for the stars that have
since formed).  \citet{bpbc11} noted an order of magnitude increase in
the molecular mass for the most luminous star forming galaxies from now
to $z \sim 0.4$, mirroring the increase in the cosmic star formation
rate density $\rho_{\rm SFR}(z)$ \citep[e.g.][]{hb06}.  They argue that
this implies that recent star formation evolution is largely due to the
run down in the available ISM supply.  \citet{hanish+06} made a similar
argument based on the slope in $\rho_{\rm SFR}(z)$ being similar to that
expected from the star formation law.  As outer disks spread and evolve,
$Q$ should remain constant with radius. The exact values of $Q$ and
$\sigma$ will be set by feedback (with less efficient star formation in the
thinning disk) and the available angular momentum.

Finally, our results do not disprove MOND, nor do they rule
out the possibility that some or most of the DM is in a gaseous form.
However, some of the appeal of these theories is that it was not clear
under the prevailing CDM paradigm why dissipative gaseous disks should
trace the non-dissipative CDM halo which resides in a spheroid.  Our work
provides this linkage by showing that ISM disks will trace DM as a
natural consequence of disk stabilisation, and our tests of the \HI\
dominated outer disks are consistent with that interpretation.

\section*{Acknowledgments}

GRM thanks Claude Carignan for first pointing out the relationship
between DM and \HI\ to him in 1992.  We thank John Cannon, Ed Elson,
Gianfranco Gentile, Christian Struve, Brad Warren, and Tobias Westmeier
for providing us with data from their studies. We are especially
grateful to Tobias Westmeier for combining his ATCA data of NGC~300 with
archival Parkes observations and extracting a new radial profile for us.
We thank Ken Freeman and Alan Duffy for useful comments, and the referee
Alessandro Romeo for making suggestions that improved the paper.  GRM
was supported in part by a Research Collaboration Award from the
University of Western Australia.  ZZ was supported through Galex GI
grant NNX09AF85G.  This research has made use of the NASA/IPAC
Extragalactic Database (NED) which is operated by the Jet Propulsion
Laboratory, California Institute of Technology, under contract with the
National Aeronautics and Space Administration.

\bigskip
\parindent=0mm {\bf Note added in proof:} We note a previous paper that reached similar
conclusions that had escaped our attention.  Struck-Marcell (1991, ApJ,
368, 348) used a related approach, the assumption that gas disks
maintain a balance of hyrdostatic forces, to derive a disk structure of
$\Sigma_g \propto R^{-1}$ for disks having a flat RC. That paper briefly
alludes to the relationship with DM, but does not generalize the problem
to arbitrary RC shape as we have done.

\bibliographystyle{mn2e}
\bibliography{mn-jour,outerq}

\end{document}


{\bf Table 2.} Fitted quantities\\
{\tiny
 
\begin{tabular}{@{}lrrrrrrrrrrrrrrr}
\hline
&&& \multicolumn{3}{c}{RC fit} & \multicolumn{3}{c}{$\Sigma_g$ fit} & \multicolumn{3}{c}{$\kappa$ fit} & \multicolumn{2}{c}{$\log(\zeta)$} & \multicolumn{2}{c}{$\log(Q)$} \\
Galaxy & $R_1$ & $R_2$ & zpt & $\gamma$ & $\epsilon$ & zpt & $N$ & $\epsilon$ & zpt & $M$ & $\epsilon$ & avg & $\epsilon$ & avg & $\epsilon$ \\
(1) & (2) & (3) & (4) & (5) & (6) & (7) & (8) & (9) & (10) & (11) & (12) & (13) & (14) & (15) & (16) \\
\hline
DDO154    &  2.4 &  5.4 & 1.42 & $ 0.417$(0.036) & 0.005 & 1.07 & $-1.16$(0.08) & 0.02 & $-1.40$ & $-0.68$(0.06) & 0.021 & $-0.96$ & 0.10 & 0.53 & 0.05 \\
NGC3741   &  1.8 &  4.5 & 1.36 & $ 0.470$(0.077) & 0.012 & 0.77 & $-0.86$(0.09) & 0.04 & $-1.33$ & $-0.66$(0.07) & 0.016 & $-1.20$ & 0.11 & 0.77 & 0.06 \\
ESO215    &  3.9 &  7.0 & 1.57 & $ 0.155$(0.005) & 0.006 & 1.52 & $-0.81$(0.15) & 0.01 & $-0.90$ & $-1.15$(0.12) & 0.005 & $-0.49$ & 0.08 & 0.07 & 0.03 \\
NGC2366   &  2.3 &  4.9 & 1.67 & $ 0.311$(0.086) & 0.010 & 1.15 & $-1.49$(0.11) & 0.03 & $-1.41$ & $-1.16$(0.09) & 0.016 & $-0.71$ & 0.08 & 0.28 & 0.04 \\
IC2574    &  4.5 &  7.5 & 1.82 & $ 0.809$(0.091) & 0.008 & 0.81 & $-0.67$(0.10) & 0.04 & $-1.77$ & $-0.42$(0.09) & 0.006 & $-0.85$ & 0.06 & 0.16 & 0.04 \\
NGC2915   &  3.6 &  7.9 & 1.94 & $-0.054$(0.109) & 0.008 & 0.84 & $-0.95$(0.10) & 0.02 & $-0.94$ & $-1.16$(0.08) & 0.017 & $-1.37$ & 0.15 & 0.85 & 0.04 \\
NGC300    &  5.3 & 12.5 & 1.96 & $ 0.137$(0.057) & 0.011 & 0.85 & $-2.01$(0.14) & 0.01 & $-1.59$ & $-1.25$(0.11) & 0.013 & $-1.10$ & 0.14 & 0.46 & 0.09 \\
ADBSJ1338 &  9.5 & 17.5 & 1.93 & $ 0.113$(0.017) & 0.007 & 1.41 & $-1.64$(0.17) & 0.05 & $-0.98$ & $-1.25$(0.14) & 0.004 & $-1.39$ & 0.06 & 0.68 & 0.05 \\
NGC7793   &  3.1 &  5.7 & 2.08 & $ 0.240$(0.042) & 0.013 & 0.71 & $-1.46$(0.10) & 0.02 & $-1.60$ & $-1.51$(0.08) & 0.020 & $-1.21$ & 0.16 & 0.46 & 0.01 \\
NGC2403   &  5.9 & 12.4 & 2.10 & $ 0.058$(0.015) & 0.004 & 0.78 & $-1.12$(0.04) & 0.03 & $-1.62$ & $-0.82$(0.03) & 0.017 & $-1.19$ & 0.09 & 0.40 & 0.05 \\
NGC3198   & 13.2 & 27.8 & 2.20 & $-0.077$(0.018) & 0.005 & 0.78 & $-0.95$(0.07) & 0.02 & $-1.80$ & $-1.03$(0.06) & 0.013 & $-0.95$ & 0.05 & 0.18 & 0.02 \\
NGC3621   &  9.1 & 18.0 & 2.15 & $ 0.002$(0.013) & 0.004 & 0.85 & $-0.32$(0.06) & 0.03 & $-1.70$ & $-0.73$(0.05) & 0.036 & $-0.99$ & 0.14 & 0.16 & 0.07 \\
NGC4736   &  2.3 &  7.3 & 2.18 & $-0.171$(0.015) & 0.018 & 0.16 & $-1.21$(0.04) & 0.14 & $-1.51$ & $-1.50$(0.03) & 0.033 & $-1.49$ & 0.55 & 1.12 & 0.17 \\
NGC6946   &  8.1 & 16.5 & 2.30 & $ 0.002$(0.007) & 0.009 & 0.80 & $-1.00$(0.05) & 0.05 & $-1.58$ & $-0.77$(0.04) & 0.037 & $-1.38$ & 0.09 & 0.40 & 0.04 \\
NGC5055   & 10.3 & 28.6 & 2.31 & $-0.091$(0.013) & 0.008 & 0.44 & $-1.54$(0.03) & 0.12 & $-1.72$ & $-1.16$(0.02) & 0.023 & $-1.51$ & 0.17 & 0.66 & 0.14 \\
NGC2903   &  9.8 & 19.7 & 2.30 & $-0.114$(0.021) & 0.004 & 0.53 & $-1.03$(0.07) & 0.06 & $-1.77$ & $-1.17$(0.06) & 0.006 & $-1.34$ & 0.07 & 0.48 & 0.06 \\
NGC3521   &  9.6 & 18.5 & 2.36 & $-0.148$(0.063) & 0.009 & 1.03 & $-1.47$(0.08) & 0.04 & $-1.58$ & $-1.28$(0.07) & 0.004 & $-1.04$ & 0.07 & 0.18 & 0.05 \\
NGC3031   &  7.8 & 12.3 & 2.32 & $-0.555$(0.031) & 0.002 & 0.84 & $-0.49$(0.10) & 0.02 & $-1.69$ & $-1.46$(0.08) & 0.013 & $-0.77$ & 0.18 & 0.30 & 0.06 \\
NGC2841   & 19.5 & 37.6 & 2.46 & $-0.126$(0.025) & 0.005 & 0.32 & $-1.78$(0.06) & 0.03 & $-1.88$ & $-1.19$(0.05) & 0.010 & $-1.59$ & 0.03 & 0.57 & 0.05 \\
NGC1167   & 24.3 & 50.4 & 2.57 & $-0.146$(0.017) & 0.019 & 0.29 & $-1.45$(0.14) & 0.05 & $-1.58$ & $-1.71$(0.12) & 0.016 & $-1.85$ & 0.27 & 0.86 & 0.08 \\
Toy model &  7.7 & 21.5 & 2.13 & $ 0.131$(0.006) & 0.005 & 0.63 & $-1.71$(0.05) & 0.03 & $-1.57$ & $-1.00$(0.04) & 0.003 & $-1.62$ & 0.07 & 0.75 & 0.09 \\
\hline
\end{tabular}} 
\\
Columns: (1) gives the galaxy name; (2) and (3) give the radii in kpc
enclosing 25\%\ and 75\%\ of the \HI\ mass respectively; columns (4)-(6)
relate to the power law fits to the rotation curves - (4) gives the
zeropoint, (5) the power-law index, and (6) the dispersion of the
residuals in dex; columns (7)-(9) relate to the power law fits to the
$\Sigma_g$ (ISM surface mass density) profiles, with the individual columns
following the pattern of (4)-(6); columns (10)-(12) relate to the power
law fits to the $\kappa$ (epicyclic frequency) profiles,  with the individual columns
following the pattern of (4)-(6); columns (13) and (14) give the average
logarithm of $\zeta$ (local gas fraction) between $R_1$ and $R_2$ and
its dispersion, respectively; columns (15) and (16) give the average
logarithm of $Q$ (disk stability parameter) between $R_1$ and $R_2$ and
its dispersion, respectively.  Values in parenthesis are the formal
uncertainties in the power law slopes from the fits to the relevant
profiles. 